\documentclass[12pt,a4paper,oneside]{article}

\usepackage[latin1]{inputenc}
\usepackage{graphicx}
\usepackage{dcolumn}
\usepackage{bm}
\usepackage{amsmath}
\usepackage{amssymb}

\title{Long range Kondo signature of a single magnetic impurity}

\author{Henning Prüser$^{1}$, Martin Wenderoth$^{1\ast}$, Piet E. Dargel$^{2}$, Alexander Weismann$^{1}$,\\
Robert Peters$^{2}$, Thomas Pruschke$^{2}$ and Rainer G. Ulbrich$^{1}$\\
\\
\normalsize{$^{1}$IV. Physikalisches Institut, Universit\"at G\"ottingen,}\\
\normalsize{Friedrich Hund Platz 1, 37077 G\"ottingen, Germany}\\
\normalsize{$^{2}$Institut für Theoretische Physik, Universit\"at G\"ottingen,}\\
\normalsize{Friedrich Hund Platz 1, 37077 G\"ottingen, Germany}\\\\
\\
\normalsize{$^\ast$ E-mail:  wenderoth@ph4.physik.uni-goettingen.de} }
\date{}

\begin{document}
\maketitle
\baselineskip18pt
\textbf{The Kondo effect, one of the oldest correlation phenomena known in condensed matter physics \cite{Hewson1993}, has regained attention due to scanning tunneling spectroscopy (STS) experiments performed on single magnetic impurities \cite{Li1998, Madhavan1998}. Despite the sub-nanometer resolution capability of local probe techniques one of the fundamental aspects of Kondo physics, its spatial extension, is still subject to discussion. Up to now all STS studies on single adsorbed atoms have shown that observable Kondo features rapidly vanish with increasing distance from the impurity \cite{Manoharan2000, Quaas2004, Zhao2005, Iancu2006a, Neel2007, Neel2008}. Here we report on a hitherto unobserved long range Kondo signature for single magnetic atoms of Fe and Co buried under a Cu(100) surface. We present a theoretical interpretation of the measured signatures using a combined approach of band structure and many-body numerical renormalization group (NRG) calculations. These are in excellent agreement with the rich spatially and spectroscopically resolved experimental data.}
\newpage
The interaction of a single magnetic impurity with the surrounding electron gas of a non-magnetic metal leads to fascinating phenomena in the low temperature limit, which are summarized by the term Kondo effect \cite{Hewson1993}. Such an impurity has a localized spin moment that interacts with the electrons of the conduction band. If the system is cooled below a characteristic temperature, the Kondo temperature $T_K$, a correlated electronic state develops and the impurity spin is screened. The most prominent fingerprint of this many body singlet state is a narrow resonance at the Fermi energy $\varepsilon_F$ in the single particle spectrum of the impurity, called Kondo or Abrikosov-Suhl resonance. The existence of this Kondo resonance has been experimentally confirmed for dense systems with high resolution photoemission electron spectroscopy and inverse photoemission \cite{Patthey1985,Ehm2007}. Due to their limited spatial resolution these measurements always probe a very large ensemble of magnetic atoms. With its capability to study local electronic properties with high spatial and energetic resolution, Scanning Tunneling Spectroscopy (STS) has paved the way to access individual impurities \cite{Li1998, Madhavan1998}.

A theoretical prediction for the local density of states (LDOS) - the key quantity measured in STS experiments - was first provided by Újsághy et al \cite{Ujsaghy2000}. According to their calculations the Kondo resonance induces strong spectroscopic signatures at the Fermi energy whose line shape is oscillatory with distance to the impurity. Since the first STS studies in 1998 \cite{Li1998, Madhavan1998} a lot of experiments on magnetic atoms and molecules on metal surfaces were performed, all revealing Kondo fingerprints \cite{Quaas2004, Zhao2005, Iancu2006a, Neel2007, Neel2008}. However, it is worth noting that all previous STS experiments on isolated ad atoms have reported that the Kondo signature rapidly vanishes within a few angstrom and no variation of the line shape occurs (for a review on ad atom Kondo systems see \cite{Ternes2009}).

In the present work we follow a novel route and investigate single isolated magnetic impurities buried below the surface with a low temperature STM operating at 6K. It has been recently shown that the anisotropy of the copper Fermi surface leads to a strong directional propagation of quasi particles, which is called electron focusing \cite{Weismann2009}. This effect gives access to individual bulk impurities in a metal that were previously assumed to be "invisible" due to charge screening. Following these observations, dilute magnetic alloys were prepared on a clean Cu(100) single crystal by adding a small amount (0.02\%) of either Co or Fe to the topmost monolayers. We have chosen Co and Fe because of their different Kondo temperature. This allows to test the universal character of the Kondo effect. 

\begin{figure}[!ht]
\centering
\includegraphics[angle=270,width=14.0cm]{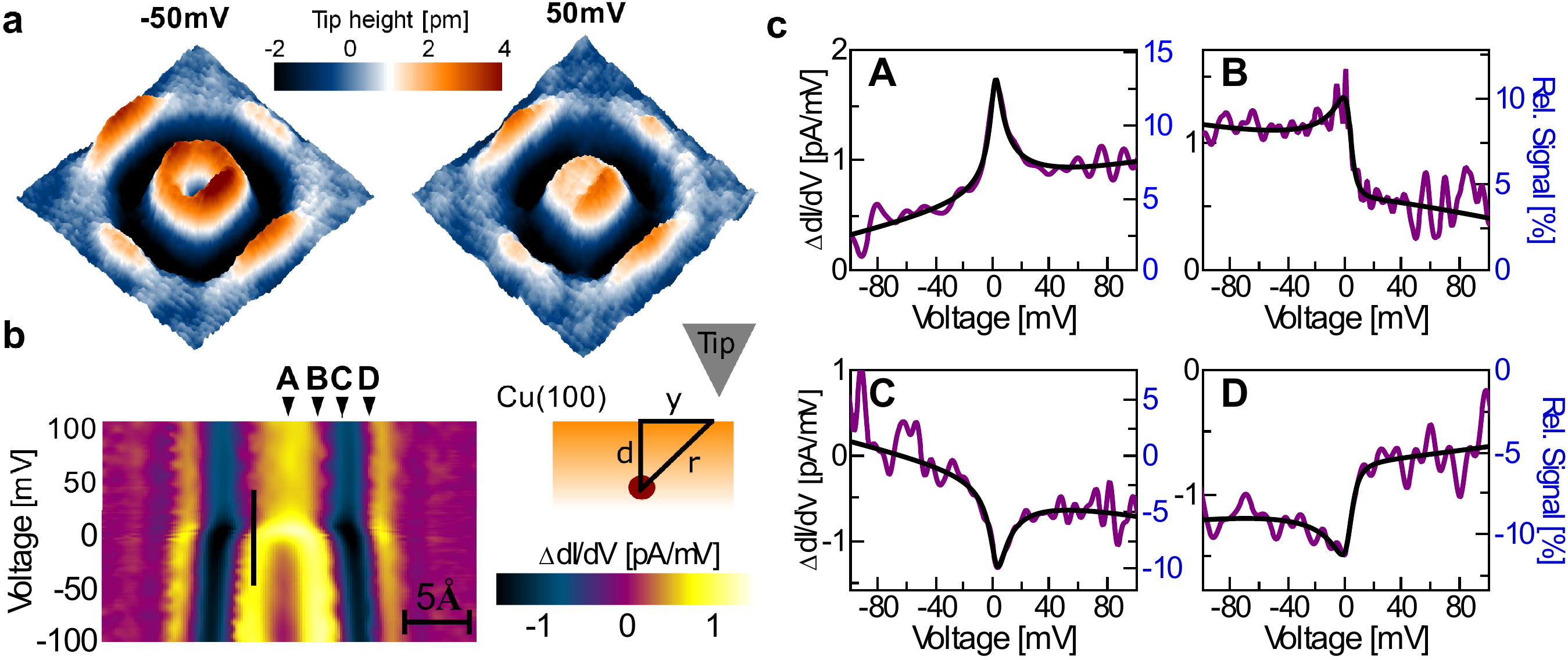}
\caption{Variation of the LDOS with the lateral distance
of the tip. a) STM topography ($1.8$nm x $1.8$nm, $2$nA) of a 4th layer Fe impurity for different bias voltages $V$. For $V < 0$ the LDOS directly above the impurity is reduced. By increasing $V$ the central ring contracts to a plateau like maximum and the size of the pattern decreases. b) Section of $\Delta \mathrm{dI}/\mathrm{dV} = \mathrm{dI}/\mathrm{dV} - \mathrm{dI_0}/\mathrm{dV}$ along the [010] direction. The tip height was adjusted to give a $\mathrm{dI_0}/\mathrm{dV}$ of $13.3$pA/mV for the free surface. An energy dependent phase shift of the interference pattern can be observed. The vertical black line would resemble the phase front for a energy independent scattering phase and indicates that the overall phase shift caused by the resonance is less than $\pi$. Single spectra (purple curve) for four different lateral distances (marked in b by black arrows) are depicted in c). To illustrate the relative amplitude of the Kondo signal, the data is normalized to the differential-conductance of the free surface (blue scale on the right side). Black curve: Calculated spectra obtained by fitting a phenomenological expression of the Kondo resonance to the $\mathrm{dI}/\mathrm{dV}$ signal (see supplementary information).} \label{fig:q_phase_lateral}
\end{figure}

As a first striking example how the Kondo effect influences the energy dependent scattering behavior on the mV scale, figure 1a shows STM topographies of a fourth layer Fe impurity for different bias voltages near zero mV. The local minimum of the LDOS present in the center of interference pattern for $V < 0$ develops into a plateau like maximum for $V > 0$. In figure 1b) the differential conductance as a function of bias voltage and one spatial coordinate $y$ across the impurity pattern is shown. The crossover observed in the topographies occurs very close to zero bias. In figure 1c) four spectra for different positions A-D are shown, illustrating very clearly that a single Kondo atom buried below the surface of copper induces long range spectral signatures that depend on the distance to the impurity. A second possibility to investigate the Kondo effect versus distance is to look at impurities situated in different depth $d$ below the surface. Figure 2 shows as an example STS-data of Co atoms, comparing single spectra (purple curves) measured directly above the impurities ($y = 0$). The lateral variations of $\mathrm{dI}/\mathrm{dV}$ are again depicted as sections (upper part of figure 2). All reveal a constriction of the pattern for positive bias voltages. The comparison between Co and Fe data demonstrates that both impurity species show similar behavior on completely different energy scales (for instance compare the 4th layer Fe in figure 1b with a 4th layer Co in figure 2).

\begin{figure}[!ht]
\centering
\includegraphics[angle=270,width=16.0cm]{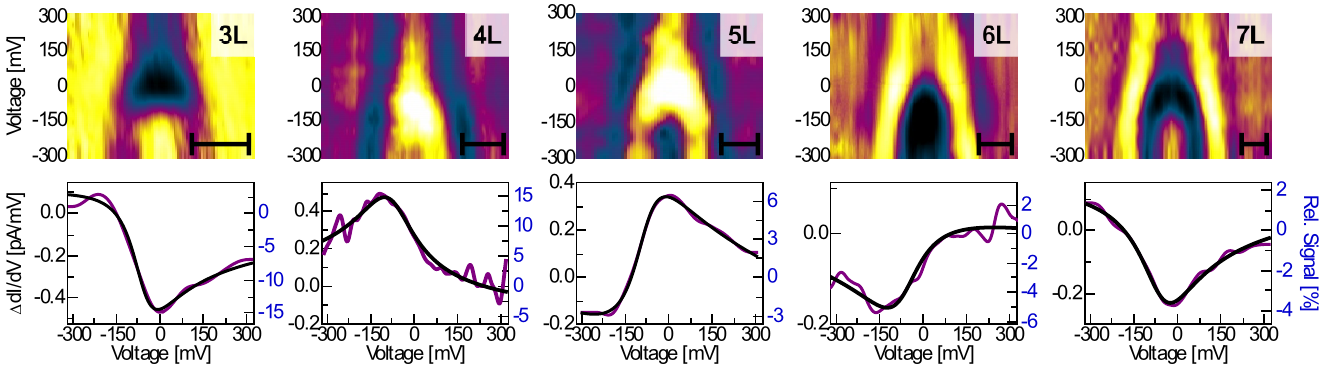}\caption{STS-data of
subsurface Co impurities in 3 to 7 layers below the Cu(100) surface. Spectrum
sections $\Delta \mathrm{dI}/\mathrm{dV} = \mathrm{dI}/\mathrm{dV} -
\mathrm{dI_0}/\mathrm{dV}$ along the [010] direction are are shown in the upper
part. A strong change in the scattering behavior can be seen for all impurity
depths. The black bar corresponds to a length of $0.5$nm for every section.
Single $\Delta \mathrm{dI}/\mathrm{dV}$ spectra (purple curve) measured with a
tip position direct above the impurity ($d = 3..7$ML) are depicted in the lower
part. The tip height was adjusted to give a differential-conductance for the
free surface $\mathrm{dI_0}/\mathrm{dV}$ of $3.2$pA/mV (3ML, 4ML, 6ML),
$5.3$pA/mV (5ML) and $6.4$pA/mV (7ML). To illustrate the relative amplitude of
the Kondo signal, the data is normalized to the differential-conductance of the
free surface (blue scale on the right side). Black curve: Calculated spectra
obtained by fitting a phenomenological expression of the Kondo resonance to the
$\mathrm{dI}/\mathrm{dV}$ signal (see supplementary information).} \label{depth}
\end{figure}

For a quantitative analysis, we have to use advanced tools of quantum many-particle theory as the Kondo problem is a genuine many-body effect. The well-known universal behavior of the Kondo effect and its associated low-energy fingerprints allows to apply the Single Impurity Anderson Model. Within this model the localized orbital of the impurity is described by (i) a single level that couples to non-interacting conduction band electrons and (ii) a Coulomb interaction $U$ between electrons on the site.

The effect of the Kondo resonance in the spectral function can be understood in analogy to other fields of physics, e.g. scattering of electrons at a potential well: the resonance causes an enhanced scattering amplitude and an energy dependent phase shift of (quasi-) particles near the resonance energy.

In many-body formalism the measured LDOS is the imaginary part of the single electron Green's function $G(\textbf{x},\textbf{x},\varepsilon)$. To calculate this quantity we use the Dyson equation, which connects the Green's function and hence the LDOS of the perturbed system to the Green's function $G^0(\textbf{x},\textbf{x}',\varepsilon)$ of the unperturbed conduction band electrons via a T-Matrix $T\left(\textbf{x}_{\text{imp}},\varepsilon\right)$. Approximating the impurity to be a point scatterer at position $\textbf{x}_{\text{imp}}$ the LDOS change is given by
\begin{eqnarray}
\Delta \text{LDOS} (\textbf{x}, \varepsilon) = -\frac{1}{\pi}\text{Im} \left[G^0(\textbf{x},\textbf{x}_{\text{imp}},\varepsilon)T\left(\textbf{x}_{\text{imp}},\varepsilon\right)G^0(\textbf{x}_{\text{imp}},\textbf{x},\varepsilon)\right]
\label{GreenDyson}
\end{eqnarray}
Calculating the band structure of copper using a linear combination of atomic orbitals (LCAO) approach (figure 3a) and treating the surface as a potential step by fixing the continuity conditions for the wave functions we obtain the free Green's functions $G^0(\textbf{x},\textbf{x}',\varepsilon)$ for copper (figure 3b and supplementary information). For the evaluation of the T-Matrix we apply the numerical renormalization group (NRG) \cite{Bulla2008}. At this point the imaginary part of the unperturbed Green's functions $G^0(\textbf{x}_{\text{imp}},\textbf{x}_{\text{imp}},\varepsilon)$ at the impurity position is needed to describe the states of the conduction band that couple to the impurity. Note that the presence of the surface leads to oscillations in this quantity (see figure 3c), which can be understood as a standing wave pattern of electrons being reflected at the surface. The NRG calculations result in a Kondo resonance, in the T-Matrix near $\varepsilon_F$ (figure 3d) with a resonance height and width showing similar oscillatory behavior as seen in the unperturbed Green's functions $G^0(\textbf{x},\textbf{x}',\varepsilon)$.

With the T-Matrix obtained by the above procedure we calculate the LDOS change using equation (1). Values for energy and hybridization of the localized orbital were taken from ab-initio results \cite{Lin2006}. The only free parameter left, the Coulomb interaction $U$, was adjusted by comparison to the experiment. The calculated LDOS sections (figure 3e) are in excellent agreement with the measured data (compare figure 2). The experimentally observed periodicity (compare a 3rd and 7th layer impurity) is also found in the simulation.

\begin{figure}[!ht]
\centering
\includegraphics[width=13.0cm]{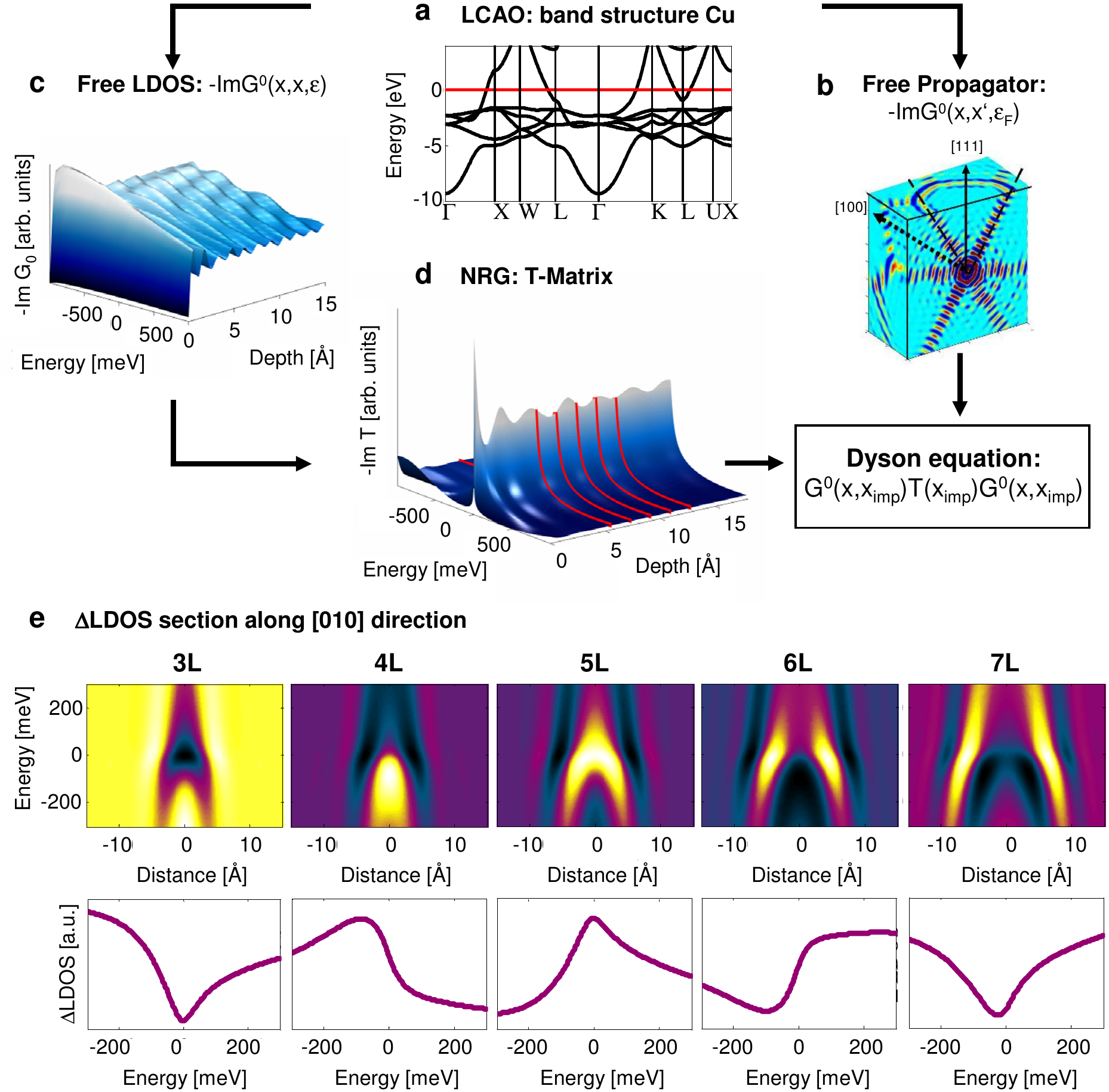}\caption{Road map of the theoretical
model: a) band structure of copper calculated using LCAO and b) the related free
propagator. c) Unperturbed LDOS as function of depth including effects of the
surface. d) The Kondo resonance for impurities located at different positions.
e) Calculated $\Delta$LDOS section along the [010] direction for Co impurities
situated in 3 - 7 monolayers below the surface. Single spectra taken directly
above the impurity position are shown below.}\label{theory}
\end{figure}

The rich spatial and spectroscopic information of the measured interference patterns allow further investigation of the Kondo resonance. The resonance width is proportional to the Kondo temperature (see supplementary information). We fit the experimental data to a phenomenological form found by Frota (see discussion further down) for all lateral positions and a constant impurity depth. Averaging over all different depths we get a Kondo temperature $T_K$ for Fe impurities of 32(16)K and for Co of 655(155)K. Results from macroscopic bulk measurements \cite{Daybell1968} give similar values.

Investigating the Kondo temperature as a function of the lateral position for a constant impurity depth we found an unexpected variation of the resonance width. As an example the 4th layer Fe impurity (figure 1c) shows a variation from $T_K\left(\mathrm{A}\right) = 29(9)$K, $T_K\left(\mathrm{B}\right) = 35(24)$K, $T_K\left(\mathrm{C}\right) = 42(18)$K and $T_K\left(\mathrm{D}\right) = 44(26)$K from position A to D. This behavior is not included in our theory so far. One possible explanation employs the actual orbital structure of the d-level in Fe or Co, which leads to a dependence of the T-Matrix on the propagation direction.

Going further, we investigate not only the long range signature of the scattering amplitude but also its phase. During our analysis it turned out that the widely used (complex) Lorentzian approximation of the Kondo resonance does not give the best description to both our experimental and theoretical data (see supplementary information). A Lorentzian for the T-Matrix results in the often used Fano line shapes \cite{Fano1961}. Better fits were obtained by a phenomenological form found by Frota \cite{Frota1986, Frota1992}. The scattering amplitude of the Kondo resonance decays much weaker with energy than a Lorentzian, which can be viewed as a minor correction. More significant, the shift of the scattering phase due to the Kondo resonance is strongly overestimated as a Lorentzian always results in a phase shift of $\pi$. In the range of energies considered here, our NRG calculations give a phase shift of only $\pi/2$ which is also descried by the phenomenological form of Frota. As one can see in figure 1b the phase shift measured in experiment is also smaller than $\pi$.

The mapping of the scattering phase as a function of energy opens a new way to discriminate between single particle resonances and Kondo scattering. Our approach allows further investigations of the properties of magnetic impurities, like the splitting of the resonance by an applied magnetic field. This was recently shown on adsorbed atoms \cite{Otte2008} and should also effect the scattering behavior of subsurface impurities. With available magnetic fields manganese may be a good candidate for such an experiment. Since we observe a long range Kondo signature this opens a new way to examine the interaction between two or more Kondo atoms with each other, or the effect of an interface in real space. Finally this observation may give an access to one of the most controversial discussed terms in Kondo physics: the meaning and size of the "Kondo cloud". With the long range LDOS signatures of buried impurities charge density oscillations are accessible and the term "Kondo cloud" may be defined in a way which is observable and consistent with experiments.

\section*{Correspondence and requests}
Correspondence and requests for materials should be addressed to Martin Wenderoth wenderoth@ph4.physik.uni-goettingen.de.
\section*{Acknowledgment}
This work was supported by the Deutsche Forschungsgemeinschaft via SFB 602
Project A3.

\section{Supplementary Information: Long range Kondo signature of a single
magnetic impurity}

\subsection{Sample preparation}
The Cu(100) single crystal substrate is cleaned by several cycles of argon
bombardment and heating. On this substrate thin films were prepared by
simultaneous deposition of copper and the impurity material from two electron
beam evaporators, leading to copper alloys with a small amount (0.02\%) of iron
or cobalt embedded. The measurements were performed with a home built low
temperature STM operating in ultra high vacuum with a base pressure better than
$5\cdot10^{-11}$ mbar. We use electrochemically etched tungsten tips, prepared
by annealing and argon ion bombardment. The performance of the tip is tuned by
controlled voltage pulses and smooth sample tip contacts.
%
All standing wave patterns are ordered to the lateral size. The depth of the
impurity is a monotone increasing function of the size of the standing wave
pattern \cite{Weismann2009}.
%
%
resolved topographies, which exists for most of the data presented in this work
(for example see figure S1). Since the Cu(100) has a fcc crystal structure and
Fe and Co atoms are situated on substitutional lattice sites, the impurity
contrast has to follow a certain ordering with respect to the host lattice. The
standing wave pattern of an odd layer impurity is centered on an corrugation
maximum of the surface. If the impurity is positioned in an even layer the
center is located between the corrugation maxima.

\subsection{STS data acquisition}
To investigate the Kondo effect and its influence on the
$\text{LDOS}\left(y,\varepsilon\right)$ we use STS spectroscopic data acquired
by recording an I(U) curve at every scanning point with interrupted feedback
loop. Further data processing included averaging and numerical differentiation.
This provides a complete map of the differential conductance
$\mathrm{dI}/\mathrm{dV}$ as a function of lateral tip position and bias voltage
on the Cu(100) surface. Since the tip is stabilized in constant current mode,
the STS data is affected by different tip-sample distances. This artifact can be
successfully removed by normalizing the STS data to a constant sample tip
distance \cite{Garleff2004}. To extract the LDOS change we subtract the
differential conductance of the free surface $\mathrm{dI}/\mathrm{dV}$. This
quantity is proportional to the change in LDOS \cite{Wahl2008} at the lateral
tip position. In order to perform a quantitative analysis of the experimental
data, drift effects which occur during acquiring the spectroscopic data, were
removed by comparing the simultaneously measured topography with a calibrated
one.

\subsection{Equation of motion technique for the SIAM}
For the simulation of the studied system we have chosen the single impurity
Anderson model (SIAM) \cite{Anderson1961}. Within the SIAM the impurity is
modeled by an effective single level with energy $\epsilon_d$. For double
occupancy the electrons are exposed to the repulsive coulomb interaction U. The
hybridization of the impurity to the copper crystal is given via the
hybridization parameter $V_k$ that describes the probability of the hopping of
an electron from a conduction electron state with momentum $k$ to the impurity
level $\epsilon_d$. Altogether the Hamiltonian for the SIAM in momentum space
reads 
\begin{eqnarray*}
H_{\textnormal{SIAM}} &=& H^0+H^{\text{Hyb}}+H^{\text{Imp}} \\
H^0&=&\sum_{k\sigma}\epsilon(k)
c^\dagger_{k\sigma}c^{\phantom{\dagger}}_{k\sigma} \\
H^{\text{Hyb}}&=& \sum_{k\sigma}\left(V^{\phantom{\dagger}}_{k}
c^\dagger_{d\sigma}c^{\phantom{\dagger}}_{k\sigma} + V^{\phantom{\dagger}}_k
c^\dagger_{k\sigma}c^{\phantom{\dagger}}_{d\sigma}\right)\\
H^{\text{Imp}}&=& \sum_\sigma \epsilon^{\phantom{\dagger}}_d
c^\dagger_{d\sigma}c^{\phantom{\dagger}}_{d\sigma} + U n_{d,\uparrow}
n_{d,\downarrow}
\end{eqnarray*}
where  $c^\dagger_{k,\sigma}$ and $c^{\phantom{\dagger}}_{k,\sigma}$ denote the
creation and annihilation operators for an electron with momentum $k$ and spin
$\sigma=\{\uparrow,\downarrow\}$ ( d denotes the d(f)-state of the impurity),
$n^{\phantom{\dagger}}_{d,\sigma}=c^\dagger_{d\sigma}c^{\phantom{\dagger}}_{
d\sigma}$ the number operator and  $\epsilon(k)$ the dispersion relation of the
free system. Setting up the equations of motion, the Green's function
$G_{k\sigma k'\sigma'}(\epsilon)$ can be written via the T-Matrix
\cite{Hewson1993} ($G^0_{k\sigma k'\sigma'}$ are diagonal in $k$ and $\sigma$)
\begin{eqnarray*}
G_{kk'}(\epsilon)=G^0_{kk'}(\epsilon)+\sum_{lm}G^0_{kl}(\epsilon)T_{lm}
(\epsilon)G^0_{mk'}(\epsilon)
\end{eqnarray*}
with the free Green's function $G^0_{k\sigma k'\sigma'}(\epsilon)=
\frac{\delta_{kk'}\delta_{\sigma\sigma'}}{\epsilon + i\eta -\epsilon(k)}$. For
the SIAM the T-matrix is given by:
\begin{eqnarray}
 T_{kk'}(\epsilon)&=& V^\ast_k G_{dd}(\epsilon) V_{k'}  \label{e:TMatrix}
\end{eqnarray}
Transforming this equation to real space yields:
\begin{eqnarray}
 G(\textbf{x},\textbf{x}',\epsilon)&=&G^0(\textbf{x},\textbf{x}',\epsilon)+\iint
d\textbf{z}' d\textbf{z}
G^0(\textbf{x},\textbf{z},\epsilon)T(\textbf{z},\textbf{z}',\epsilon)G^0(\textbf
{z}',\textbf{x}',\epsilon) \label{Dyson}
\end{eqnarray}
with
\begin{eqnarray}
T(\textbf{z},\textbf{z}',\epsilon)=
\sum_{kk'}\psi_k(\textbf{z})\psi^\ast_{k'}(\textbf{z}')T(\textbf{z},\textbf{z}',
\epsilon)
\end{eqnarray}
In the following we focus on a true point-scatterer at position
$\textbf{x}_{imp}$, leading to  
\begin{eqnarray}
G(\textbf{x},\textbf{x}',\epsilon)&=&G^0(\textbf{x},\textbf{x}',
\epsilon)+G^0(\textbf{x},\textbf{x}_{imp},\epsilon)V^2
G_{dd}(\epsilon,\textbf{x}_{imp})G^0(\textbf{x}_{imp},\textbf{x}',\epsilon)
\end{eqnarray}
In the last equation we have inserted $\textbf{x}_{imp}$ into the arguments of
$G_{dd}(\epsilon,\textbf{x}_{imp})$ to underline that the system is not truly
decoupled as 
$G_{dd}$ still depends on the free Green's function
$G^0(\textbf{x}_{imp},\textbf{x}_{imp},\epsilon)$ at the impurity position. The
free Green's functions are calculated via a Linear Combination of Atomic
Orbitals (LCAO) method including the surface and $G_{dd}(\epsilon)$ is
calculated by applying the numerical renormalization group (NRG).   

\subsection{Evaluating the free Green's function via LCAO}
The Propagator $G^0(x,x',\varepsilon)$ of the
unperturbed conduction band electrons was obtained using spectral
representation from the band structure $E_{\mathbf{k},\nu}$ of
copper and the wave functions $\Psi_{\mathbf{k},\nu}$:
\begin{equation}
G^0(\mathbf{x},\mathbf{x'},\varepsilon)=\sum_{\nu}\int_{1.BZ}d^3\mathbf{k}\frac{
\Psi_{\mathbf{k},\nu}^{\ast}(x')\Psi_{\mathbf{k},\nu}(x)}{\varepsilon-E_{\mathbf
{k},\nu}+i0^+}
\end{equation}
This quantity describes the propagation of a single electron
with energy $\varepsilon$ from a point source at $\mathbf{x'}$
to other positions $\mathbf{x}$ in the system. The band structure
$E_{\mathbf{k},\nu}$ of copper was calculated by an LCAO approach
with parameters taken from \cite{Papa1986}. The wave functions
$\Psi_{\mathbf{k},\nu}$ were derived from plane waves while
reflections of the electrons by the surface and a realistic decay
into the vacuum were taken into account. This decay depends on the energy and
the wave vector component parallel to the surface:
\begin{equation}
\kappa_{\mathbf{k},\nu}=\sqrt{\frac{2m}{\hbar^2}(\Phi-E_{\mathbf{k},\nu}
)+\mathbf{k}_{\parallel}^2}
\end{equation}
Here $\Phi=4.65 eV$ is the work function and the energy $E$ is
measured with respect to the Fermi-Level. The surface is
orientated perpendicular to the z-axis and defines the border
between the two adjacent subspaces crystal ($z\le $) and vacuum
($z > 0$). As the fcc band structure is invariant under
reflections at the (001)-plane, the wave functions are:
\begin{eqnarray}
\Psi_{\mathbf{k},\nu}(\mathbf{x}_{\parallel},z)=\left(1+\left|r_{\mathbf{k},\nu}
\right|^2\right)^{-1/2}e^{i\mathbf{k}_{\parallel}\mathbf{x}_{\parallel}}\left[e^
{ik_zz}+r_{\mathbf{k},\nu}e^{-ik_zz}\right] & \hspace{20pt} &z\le0\\
\Psi_{\mathbf{k},\nu}(\mathbf{x}_{\parallel},z)=\left(1+\left|r_{\mathbf{k},\nu}
\right|^2\right)^{-1/2}e^{i\mathbf{k}_{\parallel}\mathbf{x}_{\parallel}}
\left(1+r_{\mathbf{k},\nu}\right)e^{-\kappa_{\mathbf{k},\nu} z} & 
\hspace{20pt}&z>0
  \end{eqnarray}
with the reflection coefficient
$r_{\mathbf{k},\nu}=(ik_z+\kappa_{\mathbf{k},\nu})/(ik_z-\kappa_{\mathbf{k},\nu}
)$
providing continuity of the wave-functions and their spatial
derivative at the surface. The LDOS of the unperturbed system in
both subspaces is then:
\begin{eqnarray}
\varrho_0(z,\varepsilon)=\sum_{\nu}\int_{1.BZ}d^3\mathbf{k}\delta(\varepsilon-E_
{\mathbf{k},\nu})\left|\Psi_{\mathbf{k},\nu}(\mathbf{x}_{\parallel},z)\right|^2&
\hspace{50pt}
  \end{eqnarray}
The STM-Tip always probes the LDOS in the vacuum and the impurity
is always located within the crystal. Therefore only the Green's
function connecting a position $\mathbf{x'}=(0,0,-d)$ inside the
crystal with a position $\mathbf{x}=(x,y,h)$ in the vacuum is of
interest. Here $d$ is the depths of the impurity below the surface
and $h$ is the the tip to sample distance. This particular Green's
function is given by:
\begin{equation}
G^0(\mathbf{x}_{\parallel},h,d,\varepsilon)=\sum_{\nu}\int_{1.BZ}d^3\mathbf{k}
\frac{\left(1+r_{\mathbf{k},\nu}\right)e^{i(\mathbf{k}_{\parallel}\mathbf{x}_{
\parallel}+k_{\perp}d)-\kappa_{\mathbf{k},\nu}h}}{\left(1+\left|r_{\mathbf{k},
\nu}\right|^2\right)\left(\varepsilon-E_{\mathbf{k},\nu}+i0^+\right)}
\end{equation}
The Propagator for the opposite direction is identical due to time reversal
symmetry. Hence the vacuum LDOS at a distance $h$ above the surface which is
modified by an impurity in depth $d$ below the surface is given by:
\begin{equation}
\rho^V(\mathbf{x}_{\parallel},h,\varepsilon)=\rho_0^V(h,\varepsilon)-\frac{1}{
\pi}\mathrm{Im}\left[G^0(\mathbf{x}_{\parallel},h,d,
\varepsilon)^2T(\varepsilon)\right]
\label{finaleq}
\end{equation}
The LDOS $\rho_0^C(d,\varepsilon)$ of the unperturbed crystal
enters the NRG calculations in order to calculate the T-Matrix for the Dyson
equation (\ref{finaleq}). In the main article (see figure 3b) the vacuum LDOS
$\rho^V(\mathbf{x}_{\parallel},h,\varepsilon)$ of the perturbed system is shown
for $h=7\AA$ and $d=3...7ML$.

\subsection{Evaluating the Green's function of the impurity with the NRG}
For the evaluation of the Green's function of the impurity
$G_{dd}(\epsilon,\textbf{x}_{imp})$ we apply the Numerical Renormalization Group
(NRG). The SIAM can be interpreted as an impurity coupled to a bath of
non-interacting states. The coupling is sufficiently described by the
hybridization function $\Delta(\epsilon)$, that is proportional to the density
of conduction states that hybridize with the impurity state. For a point-like
hybridization in real space this is given by the LDOS at the impurity position
$\textbf{x}_{imp}$ and the hopping probability $V$. Therefore the hybridization
function is given by:
\begin{eqnarray}
 \Delta(\epsilon)=\pi V^2\text{LDOS}(\textbf{x}_{imp})=-V^2\mathfrak{Im}
G^0(\textbf{x}_{imp},\textbf{x}_{imp},\epsilon)
\end{eqnarray}
 The free Green's function at the impurity position is given by the LCAO
calculations. Within the NRG the hybridization function is logarithmically
discretized and mapped on a semi-infinite chain with the impurity at one end.
The logarithmic discretization ensures that the hopping amplitudes of the chain
fall off exponentially.  This allows an iterative diagonalization of the chain,
whereby at every step one site is added to the chain. 
As the increasing Hilbert space of this procedure does not allow a complete
diagonalization one has to introduce a truncation scheme in the spirit of the
renormalization group. In the NRG only the $N_s$ lowest lying eigenstates are
retained and used to build up the next Hamiltonian, thus keeping the size of the
Hilbert space constant. The result of the iterative diagonalization scheme are
the many-particle energies $E_{n}(r)$ with $r = 1,...,N_s$. For the evaluation
of the Green's function  we use the ``self-energy trick'' that maintains the
Friedel sum rule for the peak height of the Kondo resonance (for a review see
\cite{Bulla2008}). The SIAM has three free parameters $\epsilon_d,U$ and $V$.
From \cite{Ujsaghy2000, Lin2006} we take $\epsilon_d=-0.8eV$ and
$\Delta(\epsilon_F)=0.2eV$, so that hybridization is extracted accordingly to be
$V=0.447$. The interaction strength U is adjusted to the experimentally
observation. The best accordance for cobalt is given by a value of  $U=1.2eV$
and for iron for  $U=2.4eV$. On should point out that for cobalt we are not in
the true Kondo regime that is given $\frac{U}{\pi\Delta(\epsilon_F)}>3$. The
resulting spectral functions for the impurity show the Kondo resonance close the
Fermi energy. The width and height of the resonance show similar oscillatory
behavior as the hybridization function. The oscillations in the peak height are
in accordance with the Friedel sum rule $G_{dd}(\epsilon_F,\textbf{x}_{imp}) =
\frac{1}{\pi \Delta(\epsilon_F,\textbf{x}_{imp})}$. 

\subsection{Extracting the Kondo temperature}
\begin{figure}[ht]
\centering
\includegraphics[width=6cm]{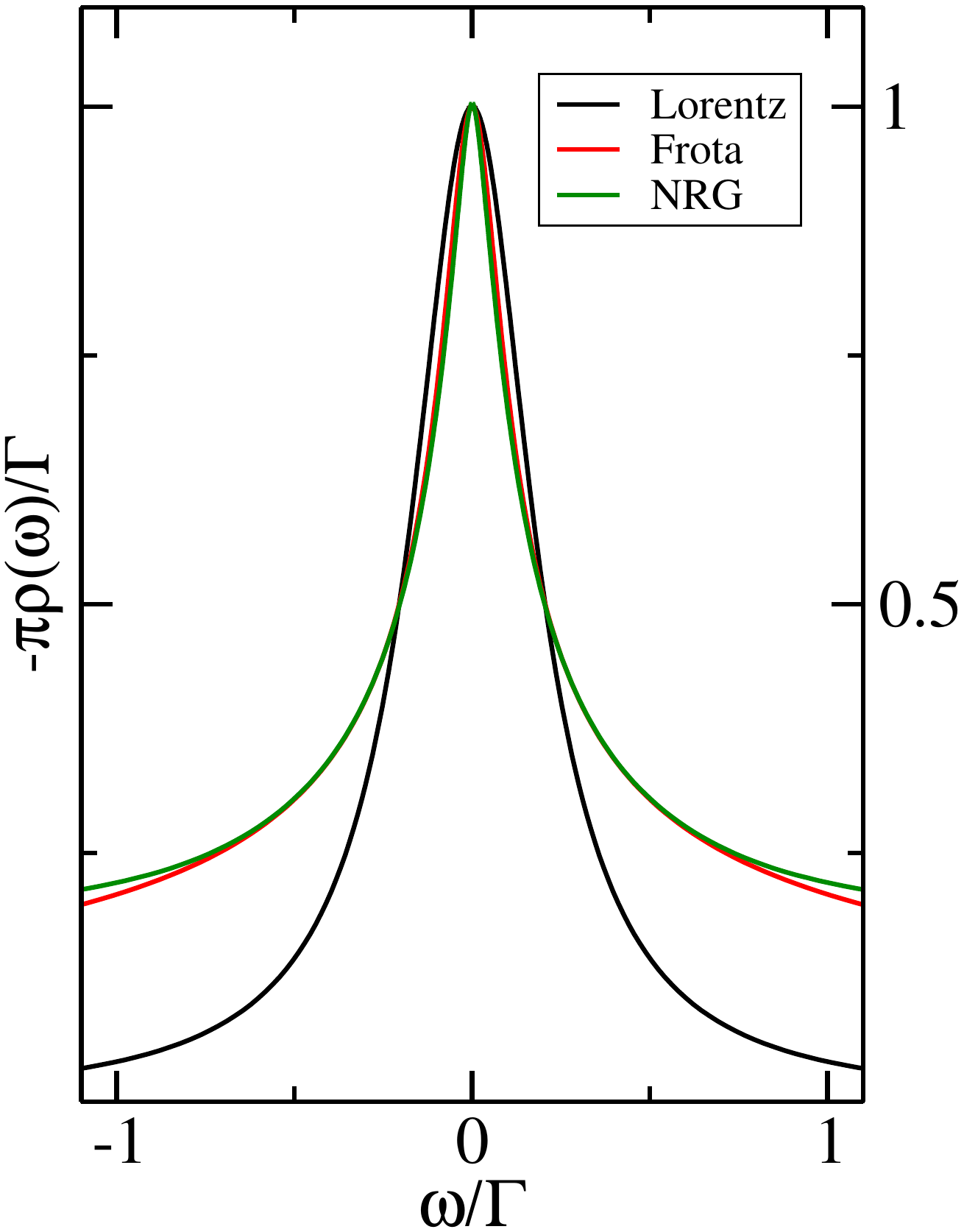}\caption{Comparison of the Kondo
resonance
calculated with the phenomenological Frota form and
Lorentzian.}\label{fig:frota}
\end{figure}

In this work we extract the Kondo temperature from the half-width at half
maximum (HWHM) of the Kondo resonance. We use Wilson's 1975 definition
\cite{Wilson1975} of the Kondo temperature $T_{KW}$:
\begin{eqnarray}
 T_{KW}&=&\bar{D}(J\rho)\exp\left[
\frac{1}{2J\rho}-\frac{1}{2}\ln|2J\rho|-1.5824|2J\rho|+O(J^2\rho^2)\right]\\
&\approx& D\sqrt2|J\rho|\exp(-1/|2J\rho|)\label{Wilson1975}
\end{eqnarray}
Extracting the Kondo Temperatures from the Kondo resonance is done via the HWHM
$\Delta_K$. In the crudest approximation the Kondo resonance is modeled via an
Lorentzian $\rho_L(\epsilon)\propto\frac{\Gamma_L}{\Gamma_L^2+\epsilon^2}$.
Actually this is only a valid approximation in the asymptotic
$\epsilon\rightarrow 0$ region \cite{Zitko2009, Bulla2000}.
The HWHM is then directly given via the $\Delta_K=\Gamma_L$.
A better line shape gives a phenomenological form found by Frota et
al.\cite{Frota1992}, \cite{Frota1986}. A comparison between the phenomenological
function found by Frota, the conventionally used Lorentzian and the resonance
calculated with the NRG are shown in figure \ref{fig:frota}. One may clearly see
the better modeling of the resonance by the Frota form.  
\begin{eqnarray}
\rho_{F}(\varepsilon)&=&\text{Im}\left[\frac{1}{\mathrm{i}}\sqrt{\frac{\mathrm{i
}\Gamma_{F}}{i\Gamma_{F}+\varepsilon}}\right]
\end{eqnarray}
The HWHM of this form is given by $\Delta_{F} = 2.542 \cdot \Gamma_{F}$.
Therefore setting this form on the same value at the HWHM as the Kondo resonance
gives $\Gamma_{F}=0.393\cdot\Delta_K $. Following \cite{Zitko2009} HWHM is
proportional to the Kondo resonance.
\begin{eqnarray}
\Delta_K^{Zitko}&<&3.7\cdot T_{KW}
\end{eqnarray}
Therefore the extracted parameter $\Gamma_{F}=0.393\cdot3.7k_BT_{KW}= 1.455k_B
T_{KW}$ is taken for the Frota form. Using this proportionality constant, 
one has to remind that it is only valid in the wide band limes for a constant
band. For the regarded system this is not the case as the presence of the
surface gives small deviations. Therefore the published values for the Kondo
Temperatures for the mean values of the resonance widths are to be interpreted
with caution.

\subsection{Phenomenological expression for the LDOS}
To extract the key quantities from the experiment and to focus on the universal
features of the Kondo resonance in the low energy regime, we use the former
mentioned phenomenological form for the T-Matrix. The free propagation of band
electrons is modeled by an energy-independent phase factor $G^0(r) \approx
e^{i\phi_q(r)/2}$ with $r = \left|\textbf{x} - \textbf{x}_{\text{imp}}\right|$.
The energy independence is justified as the energy is nearly constant around the
Fermi energy (see figure 4c). The frequency of the oscillation in the free
Green's function is given by the parameter $\phi_q(r) = 2 k_{\text{F}} \cdot r$,
which determines the line shape. Using these two approximations we finally
obtain the following fit function from equation (\ref{Dyson}) to describe the
LDOS change due to an impurity in distance $r$.
\begin{align}
    \Delta \text{LDOS} (\varepsilon, r) \propto \text{Im} \left[i
e^{i\phi_q(r)}\sqrt{  \frac{i \Gamma_F}{\varepsilon -
\varepsilon_K+i\Gamma_F}\,} \right]
\label{eqn:fit}
\end{align}
Here $\varepsilon_K$ is the position of  the Kondo resonance and
$\Gamma_F$ is proportional to the resonance width (see former section). A linear
voltage lope $\beta \epsilon + \gamma$ is added to account for additional,
approximately energy independent background scattering processes.


\end{document}